\documentclass[10pt, conference, letterpaper]{IEEEtran}
\IEEEoverridecommandlockouts
\usepackage{cite}
\usepackage{amsmath,amssymb,amsfonts}
\usepackage{algorithmic}
\usepackage{algorithm}
\usepackage{graphicx}
\usepackage{textcomp}
\usepackage{xcolor}
\usepackage{booktabs}
\usepackage{graphicx}
\def\BibTeX{{\rm B\kern-.05em{\sc i\kern-.025em b}\kern-.08em
    T\kern-.1667em\lower.7ex\hbox{E}\kern-.125emX}}
\begin{document}

\title{EdgeToll: A Blockchain-based Toll Collection System for Public Sharing of Heterogeneous Edges}

\author{\IEEEauthorblockN{
Bowen Xiao$^1$,
Xiaoyi Fan$^2$,
Sheng Gao$^3$
and Wei Cai$^1$
}
\IEEEauthorblockA{
$^1$School of Science and Engineering, The Chinese University of Hong Kong, Shenzhen, Guangdong 518172 China\\
$^2$Department of Electrical and Computer Engineering, The University of British Columbia, Vancouver V6T1Z4 Canada \\
$^3$School of Information, Central University of Finance and Economics, Beijing 100081 China
}}


\maketitle

\begin{abstract}
Edge computing is a novel paradigm designed to improve the quality of service for latency sensitive cloud applications. However, the state-of-the-art edge services are designed to specific applications, which are isolated from each other. To better improve the utilization level of edge nodes, public resource sharing among edges from distinct service providers should be encouraged economically. In this work, we employ the payment channel techniques to design and implement EdgeToll, a blockchain-based toll collection system for heterogeneous public edge sharing. Test-bed has been developed to validate the proposal and preliminary experiments has been conducted to demonstrate the time and cost efficiency of the system.
\end{abstract}

\begin{IEEEkeywords}
edge, blockchain, pricing, system, testbed
\end{IEEEkeywords}

\section{Introduction}

Cloud computing has transformed everything as a service \cite{xaas} nowadays. Nevertheless, latency sensitive cloud applications, e.g. interactive multimedia systems, are still struggling from the unacceptable delay introduced by the network round-trip time (RTT). Edge computing \cite{edgesurvey}, an emerging computing paradigm in future 5G network \cite{mobileedge5g}, is designed to improve the quality of services (QoS) for time-critical cloud applications, especially in the mobile scenarios. In contrary to the remote cloud server, edge nodes are nearby infrastructures, a.k.a. cloudlet, providing software services to the end users. On the other words, edge serves as an intermediate between a terminal device and the cloud to facilitate computing at the proximity of data sources.

However, the state-of-the-art edge platforms are specifically designed for customized applications, rather than a public service for various applications and distinct user groups. For example, an edge node deployed for power plants will not handle video processing requests from a mobile game player, even it has been staying in an idle status for a long time. The isolation among different applications significantly reduces the utilization level of edge resource, which still requires continuous maintenance work. Despite security considerations, one critical issue in preventing public edge resources sharing is the lack of motivation for the edge infrastructure provider. An incentive mechanism is still facing challenges and technical issue from a real-world implementation. First, there is no public third-party trustworthy proxy to collect toll fees for multiple edge service providers. The heterogeneous nature of edge deployments requires a transparent resource bidding platform operated independently. Second, the toll fee for a general service request is relatively small. It may be hard to use legal tender for resource pricing. Third, distinct edge platforms may adopt different pricing schemes and credit systems, which prevent the resource consumers from leveraging available edges nearby.

On the other hand, the blockchain system \cite{Nofer2017} has introduced a transparent, trustworthy and unformed ecosystem for multiple independent parties. This feature makes it a perfect solution to the toll collection problem in heterogeneous public edge sharing. The immutable and open source smart contracts \cite{smartcontractlogistics} driven by blockchain enables a transparent profit distribution scheme among multiple edge service providers in an autonomous manner. In addition, by leveraging cryptocurrency, the edge nodes from multiple service providers are able to use a unified, fine-granularity, and transparent pricing method to charge users. From the users' perspective, it is convenient to spend one cryptocurrency in consuming resources from multiple parties, which highly increase the availability of edge services. In fact, the blockchain-based toll system can minimize the cost for both providers and the users, given the business rules are well-defined: there will be no centralized operators to pocket the difference as its profit.

Nevertheless, existing blockchain systems are still in their preliminary stages. Most well-known blockchain systems are suffering from the high cost of gas fee and unacceptable latency introduce by the Proof-of-Work (PoW) \cite{Hashcash2002}, while the others, who minimize the overhead by adopting other consensus models (e.g.,  Delegated-Proof-of-Stake from EOS\footnote{https://eos.io/}), are not well recognized as full decentralized platforms. This imposes a big challenge for the toll collections systems for frequent but small amount transactions, e.g. the one we are proposing. In this work, we design and implement EdgeToll, an open source toll collection system for heterogeneous public edge sharing. By leveraging the technique of payment channel, EdgeToll provides a transparent, quick and cost-efficient solution to encourage participation of edge service providers.


The remainder of this paper is organized as follows. We reviewed related work in Section \ref{sec:related} and presented the overview of the proposed system in Section \ref{sec:overview}. We then present the technical design and test-bed implementation in Section \ref{sec:design} and Section \ref{sec:testbed}, respectively. Based on our development, the experiments are conducted to validate our system in Sections \ref{sec:experiments}. Section \ref{sec:conclusion} concludes this paper.

\section{Related Work}\label{sec:related}

\subsection{Cloud and Edge Integration}


Integrating edge to cloud platform involves a series of research topics in data and computational offloading. Traditional approach offloading schemes adopt virtualization techniques to host multiple copies of virtual machines in both cloud and edges \cite{VMCloudlet}, while another group of researchers has investigated the possibility of dynamic code partitioning \cite{CloneCloud}\cite{thinkair}. However, despite the form of offloading, the edge nodes intrinsically provide resource services for end users through direct network connectivity. In this work, we assume the end users are requesting micro-services installed in the edge nodes to simplify our model.

\subsection{Blockchain and Decentralized Applications}

A blockchain is a data structure designed to resist modifications \cite{Nofer2017}. With the help of peer-to-peer (P2P) system and proof-of-work (PoW) \cite{Hashcash2002} consensus model proposed in BitCoin \cite{bitcoin}, the decentralized ledger for cryptocurrencies became a reality. In order to add more values to the blockchain ecosystem, Ethereum \cite{ethereum} was implemented to facilitate decentralized smart contracts, which are immutable and transparent executable programs hosted by the blockchain. Nowadays, the blockchain-based decentralized applications (dApps) \cite{WeiCaiWEHFL2018} have been extended to various areas, including initial coin offerings, social networks, networked games, and IoT. In this work, we write smart contracts to develop a decentralized toll collection system for edge service sharing among multiple parties, which perfectly demonstrate the benefits of dApps.

\subsection{Payment Channel}

A payment channel \cite{LightningNetwork} is a technique designed for ``off-chain'' payments, which allows users to make multiple token transactions with a minimum number of smart contract invocations. With a typical payment channel, the payer will deposit a certain amount of tokens to the smart contract and continuously send signed micropayment to the payee without notifying the smart contract. Once the payer and the payee decided to terminate their payment process, a final signed message agreed by both parties will be posted to the smart contract, which splits the balance. In this case, only opening and closing payment channel transactions will be executed by the decentralized nodes, while an unlimited number of transactions can be performed off-chain between the participants. The state-of-the-art payment channels can be classified into two types: \textit{uni-directional payment channel} and \textit{bi-directional payment channel}. An uni-directional payment channel only allows single directional transactions, while a bi-directional payment channel \cite{Decker2015fast} allows both parties to send transactions. The duplex payment channel is composed of two uni-directional payment channels, which allows transactions to be sent from both directions. 

\section{System Overview}\label{sec:overview}

In this section, we present the overview for the proposed system.

\begin{figure}[ht]
\centering
\includegraphics[width=0.48\textwidth]{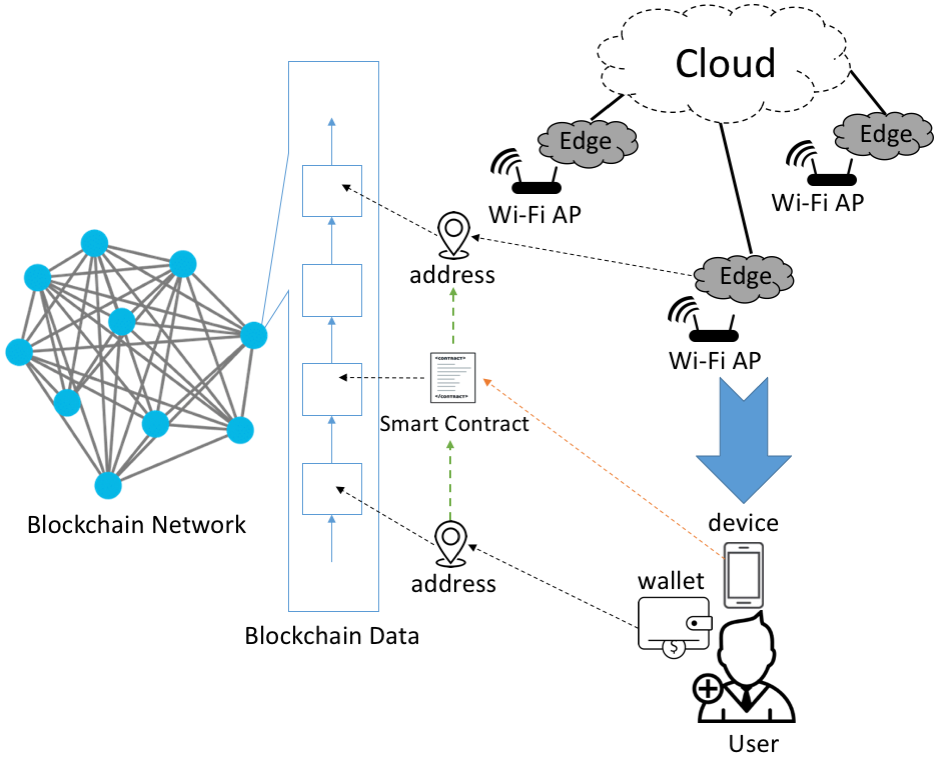}
\caption{The System Architecture for EdgeToll}
\label{fig:overview}
\end{figure}

As illustrated in Fig. \ref{fig:overview}, edge nodes, and end users should register corresponding addresses in the blockchain before they can participate in the proposed system. These addresses, being accessed with the private keys known to their owners, are the destinations of cryptocurrency tokens. From the perspective of the end user, he/she need to discover nearby edge nodes that provide the services and pay the corresponding cost after the services are delivered. In case of multiple edge nodes available, the user can choose one from these candidates, in terms of their performance and offered price. On the other hand, the edges can select their service recipients from the perspective of task complexity and bidding price, if there are multiple end users competing for the same resources. Note that, all payments should go through a smart contract to guarantee the transparency of the system.

However, the proposed system will consume significant tokens as gas fee when the users pay their toll to the edge nodes. Also, the long transaction delay will also disable high frequent service delivery to the users. Therefore, we may need to create payment channels to minimize the overhead of payment transactions. Nevertheless, it is impossible for a user to establish payment channels to a lot of edge nodes, since there will be another overhead here: the users need to lock certain among of tokens to open the channel, while he/she may only interact with one edge once.

\section{System Design}\label{sec:design}

In order to solve the above issue, we employ an open source proxy as a service matching server and the payment intermediate. The first functionality of the proxy is to match the appropriate service provider and recipient. This process can be optimized with artificial intelligence (AI) algorithms. Alternatively, this process can be a result of a series of competitions and cooperation to be modeled with game theory. In our implementation, the proxy always adopts greedy algorithms to minimize users' cost and maximize the edge nodes' profit under different scenarios. The second role of the proxy is the intermediate of users and the edges. An end user only opens one payment channel to the proxy, while the proxy open payment channels to the edges. Of course, different edge nodes from the same service provider may share the same blockchain wallet, which can significantly reduce the number of payment channels. In this work, since the payment from users to the proxy, from the proxy to the edge service providers, are uni-directional, we adopt the uni-directional payment channel.

\begin{figure}[ht]
\centering
\includegraphics[width=0.48\textwidth]{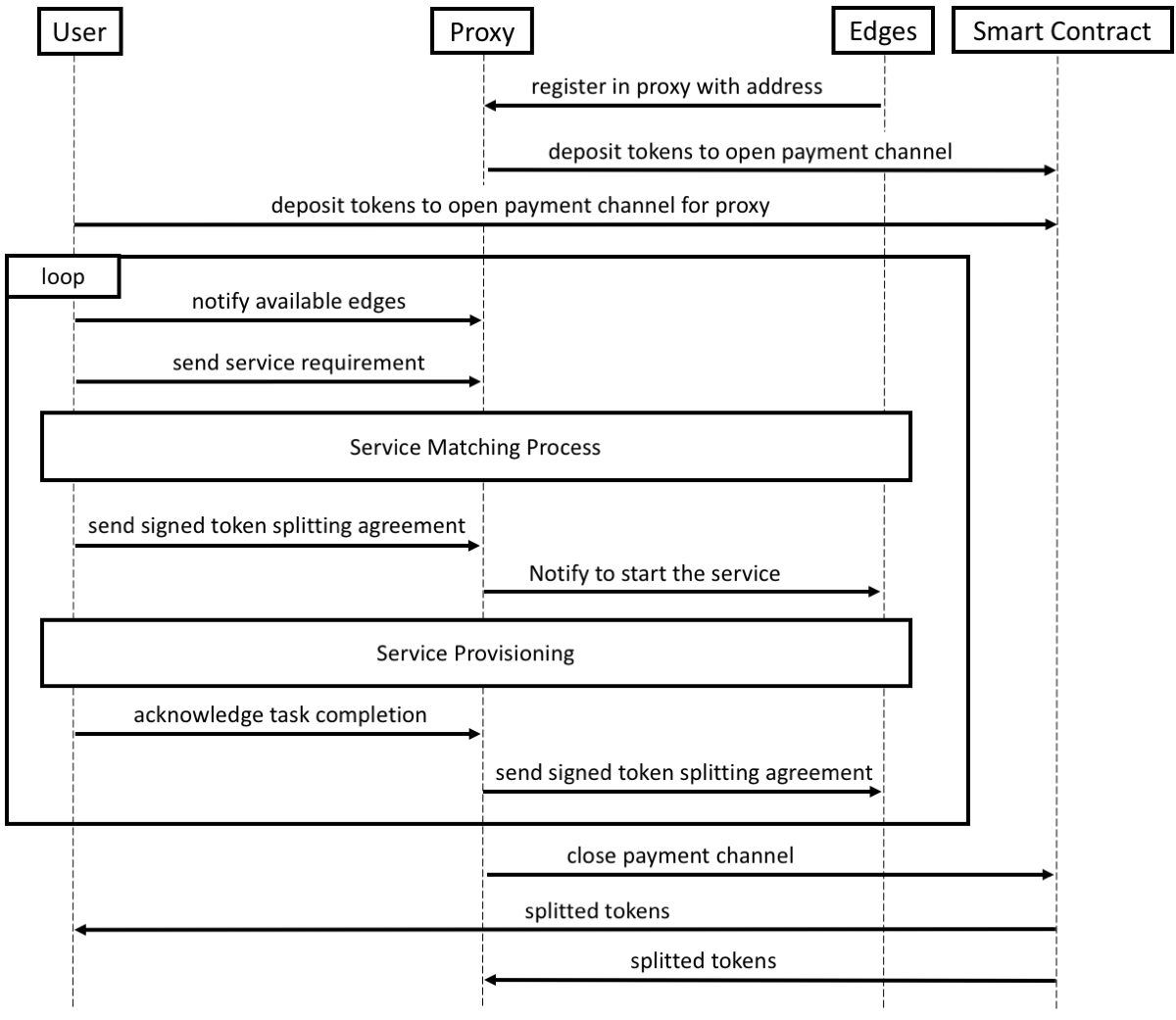}
\caption{Sequential Diagram for EdgeToll}
\label{fig:sequential}
\end{figure}

Fig. \ref{fig:sequential} illustrates the sequential diagram for the propose EdgeToll system. Edges can be deployed by any companies or individuals. For any edge who want to join in the EdgeToll public sharing platform, a registration to the proxy is required as its initialization process. Through the registration, an edge is requested to provide its blockchain address and its IP address: the former one is serving as the destination of toll fees and the later one is how the end user's device identify the edge. 

After proxy receives edge's address, the proxy will evaluate edge's condition and invoke a smart contract to open a payment channel for the registered edge. The proxy deposits tokens into that contract and set the recipient to be the registered edge so that only the edge can withdraw the token. On the other hand, the end users are usually mobile terminals whose locations are changed over time. Once a user has a demand for edge resources, he/she needs to open the payment channel for the proxy through smart contracts. At the same time, the user also needs to discover nearby edges and notify the list of available candidates to the proxy. After the service requirement is sent from the user to the proxy, service matching process will be conducted to find the suitable pairs. After that, the user needs to sign a signature on an agreement to split the tokens and send it to the proxy. The signature contains transaction information, including the recipient's address, the sender's address and the amount of payment. The proxy, the recipient of the signature, can validate the agreement with blockchain data, which is a no-cost operation, since it is a simple blockchain data reading function. After the validation, the proxy notifies the corresponding edge to deliver its service to the user. Once the user acknowledges the completion of service, the proxy will sign its token splitting agreement with the edge to deliver the edge's profit. Note that, this is another signed agreement from the proxy to the edge, which is different from the one the proxy received from the user, though the two agreements may have the same amount of tokens. In practice, the proxy may charge a small amount of transaction fee to cover its operational cost in providing service matching and payment channel intermediate service. However, the transaction fee should be written in an open source program that is agreed by both parties.

After a series of payment, the users, the proxy or the edges may choose to withdraw the tokens by closing the payment channel, which will introduce a gas fee overhead, since it is an on-chain operation. However, all payment channel based off-chain transactions, as depicted in the loop of Fig. \ref{fig:sequential}, are fast data exchange without any cost.


A debatable issue for our design is that we introduced a centralized proxy which handles payments among users and edges, which violates the decentralization spirit of the blockchain. In fact, a simple trick on software engineering can minimize the impact of this concern: the proxy is a completely open source and the proxy code will be hashed and recorded in the blockchain. Any third party can audit the proxy by comparing the hash value of the running system to the blockchain recorded data, thus, maintain the transparency of the system.

\section{Test-bed Implementation}\label{sec:testbed}

In this section, we present our implementation of an open source test-bed following the above design. 

\subsection{Enabling Software Packages}

To facilitate the development process, we adopt a series of cutting-edge software to implement constructing components for the system. We select Ethereum\footnote{https://www.ethereum.org/} as our blockchain platform, due to its popularity and maturity in technical and commercial community. In our implementation, we utilize Truffle Suite\footnote{https://truffleframework.com/} to simulate a private blockchain environment for software development and the Rinkeby Testnet\footnote{https://www.rinkeby.io/} to conduct empirical experiments.

Ethereum offers Solidity\footnote{https://github.com/ethereum/solidity}, a Turing-complete programming language for smart contract development. With solidity, we implement an uni-directional payment channel to support the transactions among users, edges and, proxy. The smart contract will be invoked by web3.py\footnote{https://github.com/ethereum/web3.py} library, which is a python\footnote{https://www.python.org/} interface for interacting with the Ethereum blockchain and ecosystem. The reason for choosing web3.py rather than the web3.js framework is that our user client program and proxy server are implemented with Python.

To integrate our EdgeToll system to an edge-terminal environment, we leverage the edge platform from Jiangxing Intelligence Inc.\footnote{http://www.jiangxingai.com/}, an edge computing start-up located in Shenzhen, China. Each Jiangxing edge node provides a Wi-Fi signal as the portal to access its AI applications, including real-time face recognition and positioning. 

To facilitate dynamical edge service discovery, we adopt pywifi\footnote{https://github.com/awkman/pywifi}, a python library to search available edge services. The list of available edge access points will be updated to the proxy in real-time. After connecting to the edge, the client will initialize a TCP/IP request through the Application Programming Interface (API) offered by Jiangxing edges to submit the user's image in base64 format, and the edge will return the location of the face in the image in a JSON file\footnote{https://www.json.org/}.

The versions of software packages are listed in Table \ref{tab:software}.

\begin{table} [htbp] 
  \renewcommand{\arraystretch}{1.1}
  \caption{Software Versions}\label{tab:software}
  \begin {center}
  \begin {tabular}
  { p{0.7in}  p{0.7in}  p{0.7in}  p{0.7in} }
  \hline
    \hline
    \textbf{Name}    & Web3.py & Truffle  & Solidity \\ \hline
    \textbf{Version} & 4.8.2   & v4.1.14 & 0.424    \\ \hline
  \hline
  \end{tabular} \label{tab:chain comparison}
  \end{center}
\end{table}

\subsection{Hardware Specifications}

Jiangxing edges used in our test-bed are Acorn RISC Machine architecture (ARM) computers with 8 GB RAM and Intel i5-7300 CPU. The edge is also equipped with a TP-LINK WDR5620 wireless access point, which adopts IEEE802.11G/802.11B standard with 1200 MB wireless rate and 2.4G/5G radio frequency.

\begin{figure}[ht]
\centering
\includegraphics[width=0.48\textwidth]{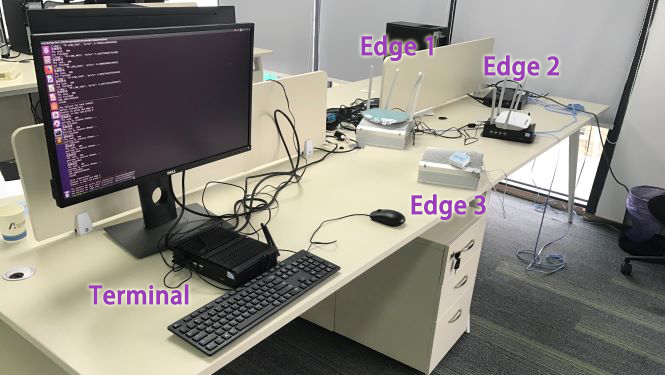}
\caption{Demonstration of the Implemented EdgeToll Test-bed}
\label{fig:demo}
\end{figure}

Fig. \ref{fig:demo} illustrates a running demonstration of our proposed system, which consists of three edge nodes and one terminal for the end user. All experiments described in following sections are conducted over the test-bed.

\section{Experiments and Analysis}\label{sec:experiments}

In this section, we validate the design and implementation of the proposed EdgeToll system with preliminary experiments.

\subsection{Experiment Design}

Because there are transaction latency and gas fee in Ethereum blockchain, overall service time and the monetary cost should be measured in our experiments. In addition, due to the resource competition among multiple users or multiple edges, the impact of the service requests frequency should be an important factor to be considered as well. Therefore, we design the following three experiments from different perspectives.

\begin{itemize}
    \item \textbf{Benefit of Payment Channel:}
    the experiment compares the time and cost efficiency with and without the utilization of payment channel technology. Our hypothesis is that with more transactions posted, payment channel will save more time and monetary cost, due to its off chain nature.
    
    \item \textbf{User Cost Minimization:}
    this experiment is designed from the perspective of end users, when there are multiple available edge nodes to use. In this scenario, the user proposes an expected price for their tasks, while proxy analyzes the status of edges to match the optimal one for the users. In fact, the edges may offer dynamic prices according to its capacity and workload, similar to the spot instance pricing\footnote{https://aws.amazon.com/ec2/spot/pricing/} available in cloud computing. According to this, we simplify our experiment by adopting the price as the only factor, thus, the proxy will match the edge with the lowest price, which is usually lower than the price proposed by the users. We denote the difference between the proposed price and the final price as the saved cost. 
    
    \item \textbf{Edge Revenue Maximization:} 
    this experiment considers the scenarios when multiple users are competing for limited edge resources, in which the users will post their expected prices together with their task requirements and the proxy may help edge to decide which user to serve first. The result may be determined by many factors, including task complexity and available resource in the edge. In this experiment we simplify the selection by adopting price users offered as the only factor, thus, to maximize the edges' own revenue. 
    
\end{itemize}

\subsection{Experimental Settings}\label{sec:experiement}

Here we present the default parameter settings for our following experiments. The default block rate in Rinkeby, approximately one block per 15 seconds, is adopted if no specific settings are imposed. The mobile terminal is a single board computer with Ubuntu 16.04 Linux operating system. By default, we iterate the numbers of users' tasks from 1 to 50 with a step of 5. With payment channel, we assume the users will not close the channel until they complete all of their tasks. Each set of experiments has been repeated for 100 times and their average values were derived as our final results.

\subsection{Result Analysis}

\begin{figure}[ht]
\centering
\includegraphics[width=0.5\textwidth]{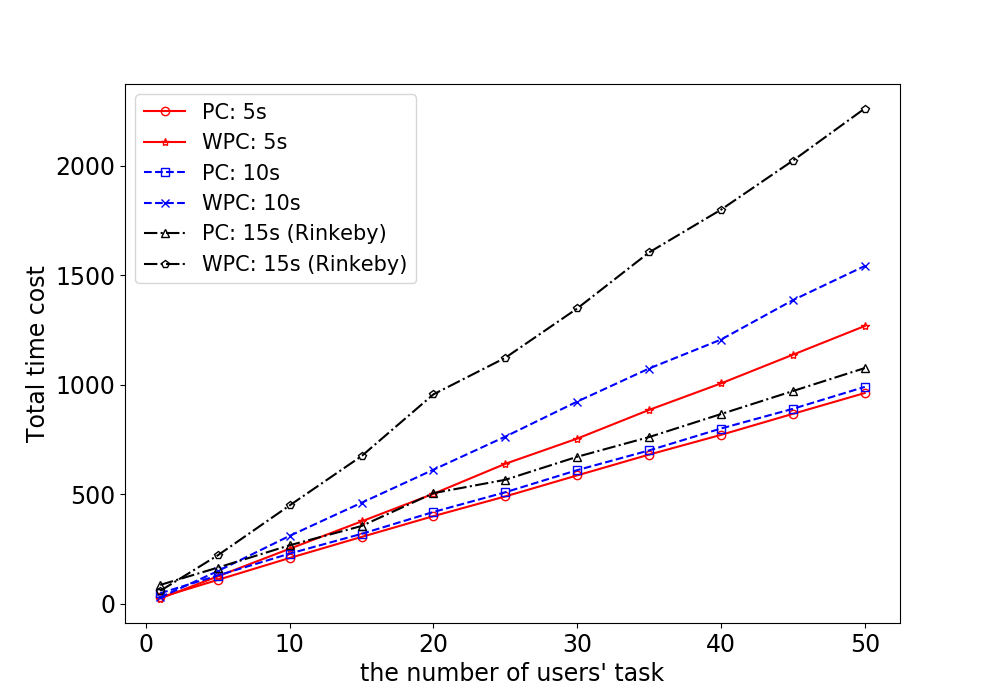}
\caption{The impact of task number on the total complete time}
\label{fig:totaltime}
\end{figure}

\begin{figure}[ht]
\centering
\includegraphics[width=0.5\textwidth]{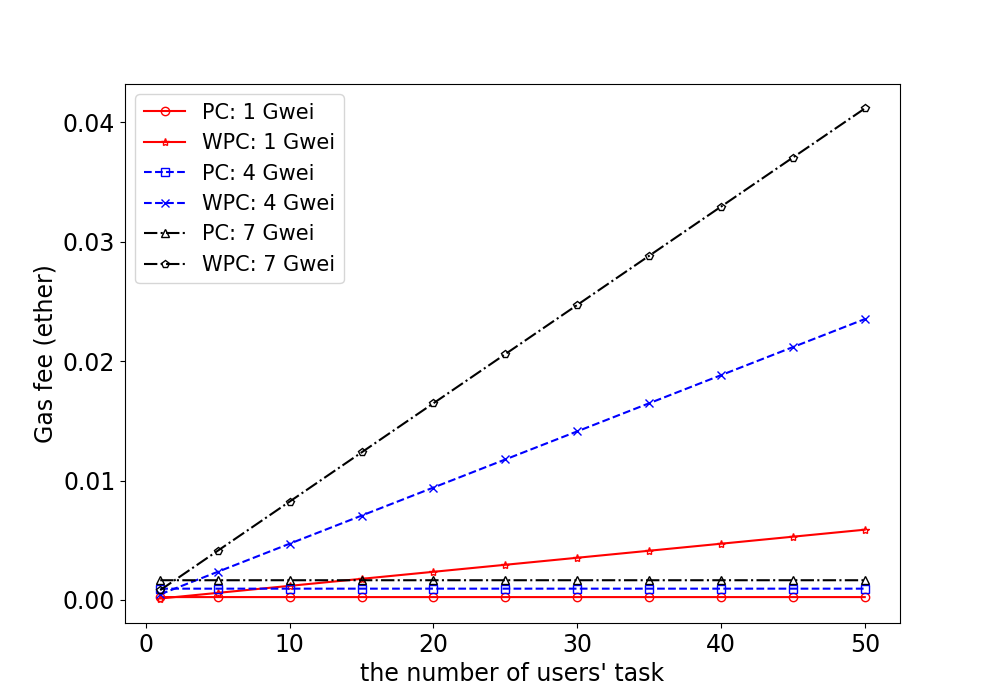}
\caption{The impact of task number on the transaction gas fee}
\label{fig:gasfee}
\end{figure}

Fig. \ref{fig:totaltime} and Fig. \ref{fig:gasfee} illustrate the performance comparisons between the system with and without the support of the payment channel. We set up different parameters as depicted in the legends, where \textit{PC} represents using the payment channel, while \textit{WPC} represents the system without using a payment channel. The value of 5s, 10s, and 15s represent the different block intervals used in Fig. \ref{fig:totaltime}, while 1 Gwei, 4 Gwei, and 7 Gwei in Fig. \ref{fig:gasfee} represent different gas fee required for posting one transaction to the blockchain.

In Fig. \ref{fig:totaltime}, we study the impact of users' task numbers on the total completion time, which consists of edge service time and the blockchain transaction delay for the toll payment when applicable. From the results, it is obvious that the total time cost increases linearly as the growth of numbers of users' task. When the total number of users' tasks is small, e.g. 1 task, the total time cost of  EdgeToll may be no different to that of systems without payment channel. However, the total waiting time values for conventional approaches, who directly pay tokens through blockchain transactions, are increased at a much higher speed, especially when the block interval is relatively high. For example, the difference of total service time cost between two schemes is 1185 seconds when the block time is set 15 seconds, a common Rinkeby scenario, which means the payment channel can reduce the time cost by more than 110\% from \textit{PC}. In fact, the largest overall latency reduction in different block time can be up to 31.8\%, 39.6\%, 110\% in the ratio of \textit{PC}, respectively.

A similar phenomenon can be observed in Fig. \ref{fig:gasfee}, which shows the difference of gas fee the system need to consume between the two paradigms. One significant feature is that, the gas fee for payment channel based experiments is a constant, no matter how many tasks are posted by the users. This is because all payments for their tasks are sent through the channel, which is a no-cost off-chain process. In fact, the only gas fees they need to pay are the opening and closing transactions in the beginning and the end of their service usage. Things are completely different without the help of payment channel: the total gas fee may increase as the total number of tasks increase. And if the price for the gas increase, the slope will become larger as well. For instance, when a user has a heavy workload e.g. 50 tasks and high gas price ( gas price = 7 Gwei), the total gas fee without payment channel will be nearly 40 times larger than that of the proposed EdgeToll system. In fact, even in the low gas price, e.g., gas price = 1 Gwei, the gas fee without payment channel can also be 6 times larger when user's workload is heavy as 50 tasks.

\begin{figure}[ht]
\centering
\includegraphics[width=0.5\textwidth]{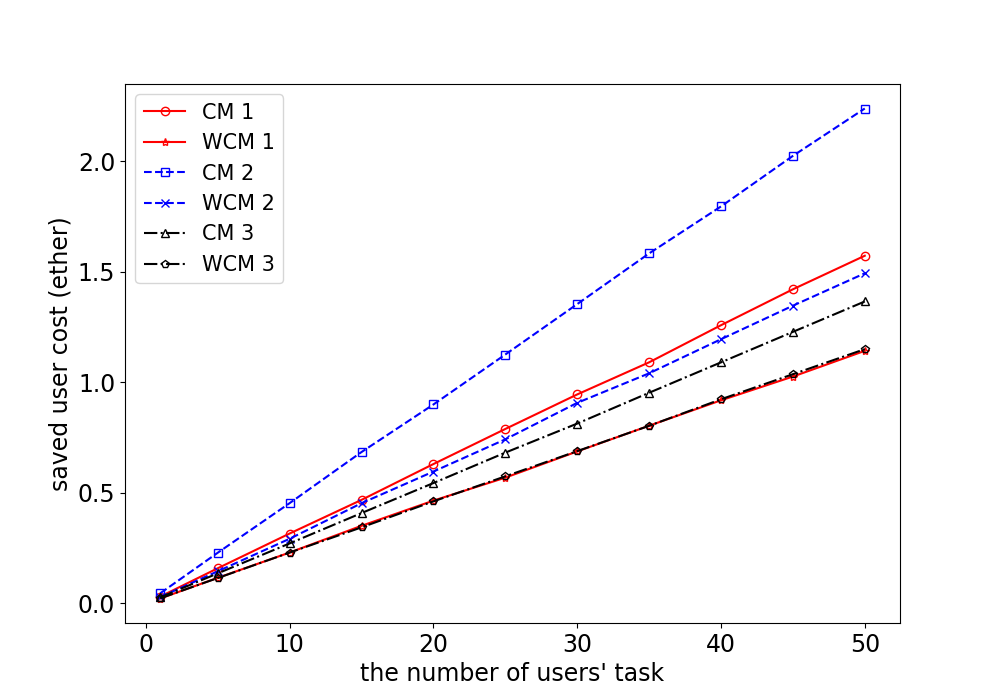}
\caption{The impact of task number on the saved cost}
\label{fig:costsaving}
\end{figure}

Fig. \ref{fig:costsaving} shows the result of user cost minimization. In the figure, \textit{CM} represents the cases using user cost minimization algorithms in the proxy, while \textit{WCM} represents the scenarios that the end users randomly select one nearby edge to post their requests. Regarding the dynamic pricing data proposed by the edges in our experiment, we are not able to find corresponding data set for a trace-drive simulation. However, we believe the spot instance price from Amazon Web Service (AWS) can be a reference for us, since they are intrinsically the same mechanism: unit prices are subject to the available resources can be provided. Therefore, we choose a number of random functions to generate dynamic prices for edge nodes, while the mean values of the normal random distribution function can be attained from observing the mean of price history in amazon web service, Linux/Unix d2.xlarge products. In Fig. \ref{fig:costsaving}, different schemes are corresponding to different random function. The numeric value \textit{1} indicates a normal distribution with mean = 0.207 and standard deviation = 0.01, the value \textit{2} implies 
a uniform distribution with interval = [0.17, 0.23], and the value of \textit{3} means another normal distribution with mean = 0.207 and standard deviation = 0.005. As shown in the figure, the system can bring remarkable benefits to the user. When the price is relatively stable, for example, when the standard deviation is 0.005, the improvement of the system is relatively insignificant. However, when the price vacillates in a uniform random function, the overall saved cost for the users with 50 tasks is around 0.73 ether, nearly 50\% reduction in comparison to a traditional system.

\begin{figure}[ht]
\centering
\includegraphics[width=0.5\textwidth]{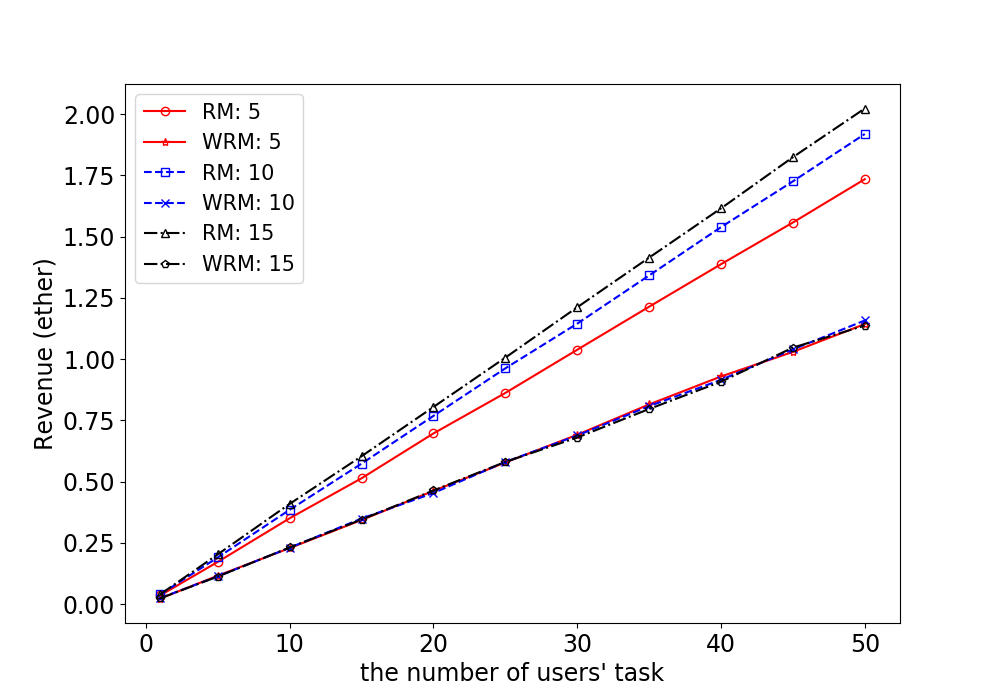}
\caption{The impact of task number on the revenue of edges}
\label{fig:revenue}
\end{figure}

Fig. \ref{fig:revenue} depicts the results of edge revenue maximization. In this figure, \textit{RM} represents the scenario that using revenue maximization of the proposed system, while \textit{WRM} represents the case that an edge will pick users' proposal in random. The numeric value after the scheme abbreviation is the number of competing users in the experiments. For example, \textit{RM:15} indicates the result of resource maximization with 15 competing users. Our hypothesis is that fiercer competition may introduce more space for the revenue maximization. To simplify our simulation, the proposed price of users is generated by normal function with mean = 0.23 and standard deviation = 0.01.

Similar to the graph measuring in user cost minimization, overall revenue derived by the edges increases along with the growth in the number of users' tasks. However, given the edges randomly select their service recipients, there will be no difference in overall revenue no matter how many users are competing for the resource. This is well illustrated by the three overlapping lines for the three \textit{WRM} schemes. In contrast, the greedy revenue maximization algorithm can bring distinct profits for the edges, given the numbers of the users are different. Given the cases of 50 tasks to be completed, the profit enhancement for 15 competitors case is about 1 ether (nearly 100 \%), while that of 5 competitors is around 0.7 ether (nearly 70 \%). This result conforms to the supply and demand rule in a free market.

\section{Conclusion}\label{sec:conclusion}

An efficient toll collection system is the key to motivate the heterogeneous edge platforms to share their vacant resource from a commercial point of view. In this work, we design and implement a blockchain-based system to fill the blank in this area. By leveraging the payment channel technique, we provide a quick and cost-efficient solution for a decentralized, transparent and trustworthy toll collection. The system is significant and powerful because it can reduce the cost of gas fee and total time, benefits both users and edges from economic perspective. Most importantly, the successful building of the system will contribute to the public popularization of edges, which can reduce the computational pressure in cloud service and accelerate the future of the Internet of Things (IoT). 

\section*{Acknowledgment}

This work was supported by the Natural Sciences and Engineering Research Council of Canada (NSERC), Nature Key Research and Development Program of China (2017YFB1400700), the National Natural Science Foundation of China (61602537, U1509214).

\bibliographystyle{ieeetr}
\bibliography{blockchain}

\end{document}